\documentclass[aps,floatfix,amsmath,nofootinbib,amssymb,superscriptaddress]{revtex4}
\usepackage{overpic}
\usepackage{amssymb}
\usepackage{indentfirst}
\usepackage{feynmf}
\usepackage{slashed}
\usepackage{cases}
\usepackage{color}
\usepackage{multirow}
\usepackage{epstopdf}
\usepackage{epstopdf}
\usepackage{amsmath}
\usepackage{float}

\usepackage[colorlinks,
citecolor=blue,
anchorcolor=red,
menucolor=red,
linkcolor=red,
filecolor=red,
runcolor=red,
urlcolor=blue,
frenchlinks=red]{hyperref}

\begin{document}
\title{Testing the Gaussian Ansatz for TMD PDFs Through CMS Measurements}
\author{Shuailiang Yang}
\affiliation{Institute of Frontier and Interdisciplinary Science, Shandong University, Qingdao, Shandong 266237, China}
\affiliation{School of Physics, Zhengzhou University, Zhengzhou, Henan 450001, China}
\author{Yingqi Hou}
\affiliation{Institute of High Energy Physics, Chinese Academy of Sciences, Beijing, 10049, China}
\affiliation{School of Physics, Zhengzhou University, Zhengzhou, Henan 450001, China}
\author{Xiaoyu Wang}
\email{xiaoyuwang@zzu.edu.cn}
\affiliation{School of Physics, Zhengzhou University, Zhengzhou, Henan 450001, China}

\begin{abstract}
We investigate the differential cross section of the $pp$ Drell-Yan process within the framework of Transverse Momentum Dependent (TMD) factorization. 
We perform phenomenological calculations of the differential cross section as a function of $q_T$ in different dilepton invariant mass bins and compare it with the measured data from the CMS experiment at $13~$TeV, employing two distinct transverse momentum dependencies of the parton distribution functions: the Collins-Soper-Sterman (CSS) evolution method and the Gaussian Ansatz. 
The results indicate that the CSS evolution method agrees well with experimental data across the entire energy range, demonstrating its strong predictive capability for the evolution behavior of TMD distribution functions. 
In contrast, the Gaussian Ansatz of the TMD PDFs is only applicable in the low-energy region and exhibits significant limitations at high energies. This study verifies the effectiveness of the TMD factorization theorem in describing high-energy processes and provides theoretical support for an in-depth exploration of the nucleon's internal structure and non-perturbative QCD dynamics.

\end{abstract}
	
\maketitle

\section{Introduction}
Transverse Momentum Dependent Parton Distribution Functions (TMD PDFs) serve as the core probes for exploring the three-dimensional internal structure of nucleons. Under the color confinement mechanism of Quantum Chromodynamics (QCD)~\cite{Gross:1973id,Politzer:1973fx,Gross:1973ju}, partons inside the nucleon not only carry longitudinal momentum but also exhibit transverse motion ($k_T$) and spatial distribution ($b$). These dynamical behaviors directly encode the non-perturbative information of QCD in the low-energy region. By introducing the transverse momentum of partons, TMD PDFs map the three-dimensional partonic structure of the nucleon onto observables, thereby serving as a crucial bridge connecting fundamental QCD theory and experimental measurements of hadron structure. Therefore, a precise understanding of the transverse momentum dependence of TMD PDFs is of indispensable significance for revealing the complete image of quarks and gluons inside the nucleon.

The theoretical foundation for describing processes sensitive to transverse momentum was established through the development of the TMD factorization theorem. The original formalism by Collins, Soper, and Sterman (CSS)~\cite{Collins:1981va,Collins:1981uk,Collins:1984kg} provided a rigorous framework to factorize cross sections into hard functions, TMD PDFs, and soft factors, while systematically resumming large logarithmic terms via the Collins‑Soper evolution equation. This formalism was later refined and extended, with key contributions from Collins in Ref.~\cite{Collins:2011zzd} and from Ji, Ma, and Yuan (JMY)~\cite{Ji:2004wu}. These modern formulations, often expressed in Fourier space with respect to the impact parameter $b$, introduced additional scale dependencies—the renormalization scale $\mu$ and the Collins‑Soper scale $\zeta$‑to properly handle rapidity divergences. The resulting factorization theorems are equivalent to those derived in Soft‑Collinear Effective Theory (SCET)~\cite{Bauer:2000ew,Bauer:2000yr,Bauer:2001ct,Bauer:2001yt} and form the cornerstone of modern TMD physics, enabling precise predictions for a wide range of processes including Semi‑Inclusive Deep Inelastic Scattering (SIDIS)~\cite{Ji:2004wu,Aybat:2011zv,Collins:2012uy,Ji:2002aa,Echevarr_a_2013,Echevarria:2012pw}, Drell‑Yan (DY) dilepton production~\cite{Boer:2006eq,Arnold:2008kf,Collins:2011zzd}, and $e^+e^-$ annihilation~\cite{Pitonyak:2013dsu,Boer:2008fr} processes.

In early phenomenological studies, the transverse momentum dependence of TMDs was often modeled using a simple Gaussian Ansatz, assuming a factorized form as $f(x, \bm{k}_T) = f(x) \frac{e^{-k_T^2 / \langle k_T^2 \rangle}}{\pi \langle k_T^2 \rangle}$~\cite{Gardiner:1970wy}. This Generalized Parton Model (GPM)~\cite{Anselmino:2005nn,Schweitzer:2010tt,Anselmino:2013lza} successfully described a wide range of low‑energy data from experiments such as HERMES and COMPASS~\cite{Cammarota:2020qcw,Schweitzer:2010tt,Anselmino:2013lza}. In this approach, the transverse momentum dependence of the cross section simplifies to a Gaussian form, where the observed transverse momentum $P_{hT}$ of the final-state hadron in SIDIS process is governed by the sum of the widths from the TMD PDF and the fragmentation function. While this simple model has the advantage of being analytically tractable and providing a satisfactory description of data within specific kinematic cuts, it violates the full QCD evolution of TMDs. Modern TMD formalism predicts a more intricate behavior in which the transverse momentum distribution is controlled by the Sudakov-like form factor that resums large logarithms and governs the scale dependence of TMDs. Consequently, the intrinsic transverse momentum distribution is no longer a fixed Gaussian but evolves with the hard scale, reflecting the underlying QCD dynamics.

Recently, the CMS Collaboration reported measurements of the double differential cross section for $pp$ Drell-Yan process as functions of $q_T$ and the invariant mass of the final-state dilepton at $\sqrt{s}=13~\text{TeV}$ with an integrated luminosity of $36.3\ \text{fb}^{-1}$~\cite{CMS:2022ubq}. 
It provides an ideal opportunity to assess how well these two distinct approaches—the simplified Gaussian parametrization and the full Sudakov evolution—describe the data across different kinematic regimes. 
In this work, we study the transverse-momentum spectrum of Drell-Yan dilepton within the TMD factorization framework. Two approaches to TMD evolution in the initial-state proton are considered, and the resulting predictions are compared with CMS data. This comparison allows us to probe the partonic structure of the proton, shedding light on its three-dimensional structure and nonperturbative QCD dynamics.

The remainder of this paper is organized as follows. In Sec.~\ref{Sec.two}, we present the theoretical calculation of the differential cross section for the Drell–Yan process in proton–proton collisions within the framework of TMD factorization. To incorporate TMD evolution effects, two approaches are employed: the CSS evolution formalism and the Gaussian Ansatz. In Sec.~\ref{Sec.there}, we provide numerical results, including theoretical predictions from both approaches and compare them with experimental data from the CMS experiment. Finally, Sec.~\ref{Sec.four} summarizes the main findings of this work and presents our conclusions.

\section{Framework}
\label{Sec.two}
The process under study is the $pp$ Drell-Yan process
\begin{align}
p\left(P_1\right)+p\left(P_2\right) \rightarrow \gamma^{\star}(q) + X \rightarrow l(\ell)+l^{\prime}(\ell^\prime)+X\left(P_X\right),
\label{eq.DYprocess}
\end{align}
where $P_1$ and $P_2$ denote the four-momenta of the two incident protons, and $q$ denotes the momentum of the virtual photon. In contrast to that in the SIDIS process, $q$ is a time-like vector in the Drell-Yan process, namely, $Q^2=q^2=m_{ll}^2>0$, which is the invariant mass squared of the final-state lepton pair. In order to express the differential cross section of the process, we introduce the following kinematical variables
\begin{align}
s=\left(P_1+P_2\right)^2, \quad  & \tau=\frac{Q^2}{s}, \nonumber\\
x_{1}=\frac{Q^2}{2 P_1 \cdot q}=\sqrt{\tau} e^{+y}, \quad  & x_{2}=\frac{Q^2}{2 P_2 \cdot q}=\sqrt{\tau} e^{-y}, \nonumber\\
x_F=2 q_L / \mathrm{s}=x_{1}-x_{2}, \quad   & y=\frac{1}{2} \ln \frac{q^{+}}{q^{-}}=\frac{1}{2} \ln \frac{x_{p 1}}{x_{p 2}},
\end{align}
where $s$ is the center-of-mass energy squared; $x_{1}$ and $x_{2}$ are the light-front momentum fractions carried by the annihilating quarks in the two initial-state protons, respectively;
$q_L$ is the longitudinal momentum of the virtual photon in the c.m. frame of the incident protons; $x_F$ is the Feynman $x$ variable, which corresponds to the longitudinal momentum fraction carried by the lepton pair; and $y$ is the rapidity of the lepton pair. Thus, the differential
cross section for the $pp$ Drell-Yan process described in Eq.~(\ref{eq.DYprocess}) has the form~\cite{Collins:1984kg}
\begin{align}
\frac{d^4 \sigma}{d Q^2 d y d^2 \boldsymbol{q_{T}}}=\sigma_0 \sum_q e_q^2 \int d^2\boldsymbol{k_{T1}}d^2 \boldsymbol{k_{T2}}f_{1,q/p_1}(x_1,\boldsymbol{k_{T1}})f_{1,\bar{q}/p_2}(x_2,\boldsymbol{k_{T2}})\delta^2(\boldsymbol{q_{T}}-\boldsymbol{k_{T1}}-\boldsymbol{k_{T2}}),
\label{eq.CrossSection}
\end{align}
where $\sigma_0=\frac{4 \pi \alpha_{\rm{em}}^2}{3 N_{C} s Q^2}$ is the cross section at tree level. $\boldsymbol{q_{T}}$ represents the transverse momentum of the virtual photon, while $\boldsymbol{k_{T1}}$ and $\boldsymbol{k_{T2}}$ denote the transverse momenta of the interacting quarks in proton 1 and proton 2, respectively. Performing the Fourier Transformation on the $\delta$-function in Eq.~(\ref{eq.CrossSection}), one can obtain the following form
\begin{align}
\frac{d^4 \sigma}{d Q^2 d y d^2 \boldsymbol{q_{T}}}&=\sigma_0 \sum_q e_q^2 \int d^2\boldsymbol{k_{T1}}d^2 \boldsymbol{k_{T2}} \int \frac{d^{2}\boldsymbol{b}}{(2 \pi)^{2}} e^{i\left(\boldsymbol{q_{T}}-\boldsymbol{k_{T1}}-\boldsymbol{k_{T2}}\right) \cdot \boldsymbol{b}} f_{1,q/p_1}(x_1,\boldsymbol{k_{T1}})f_{1,\bar{q}/p_2}(x_2,\boldsymbol{k_{T2}}) \nonumber \\
&=\sigma_0 \sum_q e_q^2 \int \frac{d^{2} \boldsymbol{b}}{(2 \pi)^{2}} e^{i \boldsymbol{q_{T}} \cdot \boldsymbol{b}} \int d^{2} \boldsymbol{k}_{T1} e^{-i \boldsymbol{k}_{T1} \cdot \boldsymbol{b}} f_{1,q/p_1}\left(x_1,\boldsymbol{k_{T1}}\right) \int d^{2} \boldsymbol{k}_{T2} e^{-i \boldsymbol{k}_{T2} \cdot \boldsymbol{b}} f_{1,\bar{q}/p_2}\left(x_2,\boldsymbol{k_{T2}}\right).
\label{eq.CrossSection_TMD}
\end{align}
Here, we neglect the energy dependence of the TMD PDFs, which will be discussed in detail in the next subsection.

\subsection{The differential cross section of the Drell-Yan process using the CSS method}
Since it is much more convenient to solve the TMD evolution equations in coordinate space ($b$-space) than in transverse momentum space ($q_{T}$), where $b$ is conjugate to $q_{T}$ via Fourier Transformation, the differential cross section is usually expressed as a $b$-dependent
function formulated according to TMD factorization. Then, it can be transformed back to transverse momentum space through inverse Fourier Transformation. The transformation of TMD PDFs from transverse momentum space to $b$-space is given by 
\begin{align}
\tilde{f}(x, b) &=\int d^{2} \boldsymbol{k}_{T} e^{-i \boldsymbol{k}_{T} \cdot \boldsymbol{b}} f\left(x,\boldsymbol{k_{T}}\right).
\end{align}
Therefore, the differential cross section in Eq.~(\ref{eq.CrossSection_TMD}) can be rewritten as
\begin{align}
\frac{d^4 \sigma}{d Q^2 d y d^2 \boldsymbol{q_{T}}}=\sigma_0 \int \frac{d^{2} \boldsymbol{b}}{(2 \pi)^{2}} e^{i \boldsymbol{q_{T}} \cdot \boldsymbol{b}} \widetilde{W}_{U U}(Q ; b)+Y_{U U}\left(Q, q_{\perp}\right).
\label{eq.CrossSection_B}
\end{align}
The structure function $\widetilde{W}_{U U}(Q ; b)$ contains all-order resummation results and dominates at low $q_{T} \ll Q$ values, while the $Y_{U U}$ term provides the necessary correction at moderate $q_{T} \sim Q$ values. In this work, we will assume $q_{T} \ll Q$, neglecting the $Y$-term, and only consider the first term on the r.h.s. of Eq.~(\ref{eq.CrossSection_B}).

In general, TMD factorization aims to separate well defined TMD functions so that they can be used in different processes while including the scheme/process dependence in the corresponding hard factors. $\widetilde{W}_{U U}(Q ; b)$ can be expressed as the convolution of the TMD distribution functions at $b$-space and hard factor~\cite{Prokudin_2015}
\begin{align}
\widetilde{W}_{U U}(Q ; b)=H_{U U}(Q ; \mu)\sum_{q, \bar{q}} e_q^2 \tilde{f}^{\rm{sub}}_{1,q / p_1}\left(x_1, b ; \mu, \zeta_F\right) \tilde{f}^{\rm{sub}}_{1, \bar{q}/p_2}\left(x_2, b ; \mu, \zeta_F\right),
\label{eq.WUU}
\end{align}
where $\tilde{f}^{\rm{sub}}_{1,q/ H}$ is the subtracted distribution function in $b$-space with the superscript “sub” representing that the soft factor is subtracted. $H_{U U}(Q ; \mu)$ is the factor associated with hard scattering, $\mu$ is the renormalization scale in the case of collinear PDFs, and $\zeta_F$ denotes an energy scale related to the cutoff of the TMD distributions. We should note that the distribution function $\tilde{f}^{\rm{sub}}_{1,q/ H}$ is universal.
However, the way to subtract the soft factor in the distribution function depends on the scheme used to regulate the light-cone singularity in the TMD definition~\cite{Collins:1981uk}, whereas the hard factor $H_{U U}(Q ; \mu)$ is also scheme-dependent. In the literature, two different schemes are
usually used: the Collins-11 scheme~\cite{Collins:2011zzd} and the Ji-Ma-Yuan scheme~\cite{Ji:2004xq}. If we perform a Fourier Transformation on $\tilde{f}^{\rm{sub}}_{1,q/ H}\left(x, b ; \mu, \zeta_F\right)$, we obtain the distribution function in the transverse momentum space, which contains information about the
probability of finding a quark with the collinear momentum fraction $x$ and transverse momentum $k_{T}$.

The TMD evolution for the $\zeta_F$ dependence of TMD PDFs is encoded in a Collins-Soper (CS)~\cite{Collins:2011zzd} equation through
\begin{align}
\frac{\partial \ln \tilde{F}\left(x, b ; \mu, \zeta_{F}\right)}{\partial \ln \sqrt{\zeta_{F}}}=\frac{\partial \ln \tilde{D}\left(z, b ; \mu, \zeta_{D}\right)}{\partial \ln \sqrt{\zeta_{D}}}=\tilde{K}(b ; \mu),
\end{align}
while the $\mu$ dependence is derived from the renormalization group equation as
\begin{align}
&\frac{d \tilde{K}}{d \ln \mu}=-\gamma_{K}\left(\alpha_{s}(\mu)\right) ,\\
&\frac{d \ln \tilde{F}\left(x, b ; \mu, \zeta_{F}\right)}{d \ln \mu}=\gamma_{F}\left(\alpha_{s}(\mu) ; \frac{\zeta_{F}^{2}}{\mu^{2}}\right), 
\end{align}	
where $\tilde{K}$ is the CS evolution kernel, and $\gamma_{K}$ and $\gamma_{F}$ are the anomalous dimensions. The solutions of these evolution equations are studied in detail in Ref.~\cite{Collins:2011zzd}. Here, we will only discuss the final result. The overall solution structure for $\tilde{f}(x,b;\mu, \zeta_F)$ is the same as that for the Sudakov form factor. Namely, the energy evolution of TMDs from an initial energy $\mu$ to another energy $Q$ is encoded in the Sudakov-like form factor $S$ through the exponential form exp($-S$). Hereafter, we set $\mu=\sqrt{\zeta_F}=Q$ and express $\tilde{f}(x,b;\mu = Q, \zeta_F = Q^{2})$ as $\tilde{f}(x,b;Q)$ for simplicity
\begin{align}
\tilde{f}_{q/H}(x,b;Q)=\mathcal{F}\times e^{-S}\times \tilde{f}_{q/H}(x,b;\mu_{B}),
\label{eq.fqh}
\end{align}
where $\mathcal{F}$ is the hard factor and depends on the scheme that we choose.

The $b$-dependence of TMD distributions determines the transverse momentum dependence of experimental observables through Fourier Transformation, which makes understanding the $b$-dependence important. In the small $b$ region, the dependence is perturbatively calculable, while at large $b$ the dependence becomes nonperturbative and should be obtained from experimental data. To combine the perturbative information at small $b$ with the nonperturbative part at large $b$, a matching procedure must be introduced with a parameter $b_{\mathrm{max}}$ serving as the boundary between the perturbative and nonperturbative regions.
A $b$-dependent function $b_\ast$ is defined, which has the property $b_\ast\approx b$ at low values of $b$ and $b_{\ast}\approx b_{\mathrm{max}}$ at large $b$ values. A typical value of $b_{\mathrm{max}}$ is chosen around 1 GeV$^{-1}$, such that $b_\ast$ is always in the perturbative region. There are several different $b_\ast$ prescriptions in the literature \cite{Collins_2016,Bacchetta_2017}. In this work, we adopt the original prescription introduced in Ref.~\cite{Collins:2011zzd}, with $b_\ast=b/\sqrt{1+b^2/b_{\rm max}^2}$.

In the small $b$ region $1/Q \ll b\ll1/\Lambda_{\textrm{QCD}}$, the defined TMDs at fixed energy $\mu$ can be expressed as a convolution of perturbatively calculable hard coefficients and the corresponding collinear PDFs~\cite{Collins:1981uk,Bacchetta_2013}
\begin{align}
\tilde{f}_{q/H}(x, b ; \mu_{B})= \sum_{i}C_{q \leftarrow i} \otimes f_{i/H}(x, \mu_{B}),
\label{eq.collinear}
\end{align}
where $\otimes$ stands for the convolution in the momentum fraction $x$
\begin{align}
C_{q\leftarrow i}\otimes f_{i/H}(x,\mu_B)& \equiv \int_{x}^{1} \frac{d \xi}{\xi} C_{q \leftarrow i}(\frac{x}{\xi},  \mu_{B}) f_{i/H}(\xi, \mu_{B}), 
\end{align}
and $f_{i/H}(\xi, \mu_{B})$ is the corresponding collinear PDF of the $i$ flavor in hadron $H$ at the energy scale $\mu$, which can be a dynamic scale related to $b_\ast$ by $\mu_B=c_0/b_*$, with $c_0=2e^{-\gamma_E}$ and $\gamma_{E} \approx 0.577$ being the Euler Constant~\cite{Collins:1981uk}. In addition, the sum $\sum_{i}$ runs over all parton flavors.
Independent of the type of initial hadrons, the perturbative hard coefficients have been calculated for the parton-target case~\cite{Aybat_2011,Collins:1981uw} and the results have been presented in Ref.~\cite{Bacchetta_2013} (see also appendix A of Ref.~\cite{Aybat_2011}) as 
\begin{align}
& C_{q \leftarrow q^{\prime}}(x, b)=\delta_{q q^{\prime}}\left[\delta(1-x)+\frac{\alpha_s}{\pi}\left(\frac{C_F}{2}(1-x)\right)\right], \label{eq.c}\\
& C_{q \leftarrow g}(x, b)=\frac{\alpha_s}{\pi} T_R x(1-x),
\end{align}
where $C_{F}=\left(N_{C}^{2}-1\right) /\left(2 N_{C}\right)={4\over3}$, $T_{R}=\frac{1}{2}$, and $\alpha_s$ is the strong coupling constant at the energy scale $\mu_{B}$. The $b_\ast$ prescription prevents $\alpha_s(\mu_{B})$ from hitting the so-called Landau
pole in the large $b$ regime. We note that the $C$-coefficients do not depend on the TMD scheme and on the initial hadrons. Namely, the $C$-coefficients in Eq.~(\ref{eq.c}) are universal for different schemes and initial hadrons.

The Sudakov-like form factor in Eq.~(\ref{eq.fqh}) can be separated into a perturbatively calculable part and a nonperturbative part
\begin{align}
{S}(Q;b)= {S}_{\rm pert}(Q;b_*)+S_{\rm NP}(Q;b).
\end{align} 
The perturbative part of $S$ has the form
\begin{align}
{S}_{\rm pert}(Q;b_*)=\int^{Q^2}_{\mu_b^2}\frac{d\bar{\mu}^2}{\bar{\mu}^2}\left[A(\alpha_s(\bar{\mu}))
\ln(\frac{Q^2}{\bar{\mu}^2})+B(\alpha_s(\bar{\mu}))\right],
\label{eq.Spert}
\end{align} 
The coefficients $A$ and $B$ in Eq.~(\ref{eq.Spert}) can be expanded as $\alpha_{S}/\pi$ series:
\begin{align}
A=\sum_{n=1}^{\infty}A^{(n)}(\frac{\alpha_s}{\pi})^n,\\
B=\sum_{n=1}^{\infty}B^{(n)}(\frac{\alpha_s}{\pi})^n.
\end{align}
In this work, we take $A^{(n)}$ to $A^{(2)}$ and $B^{(n)}$ to $B^{(1)}$ up to NLL order (with $N_{C}=3$ and $n_{f}=5$) \cite{Aybat_2011,Echevarr_a_2013,Collins:1984kg,Landry:2002ix,Qiu:2000ga,Kang:2011mr}:
\begin{align}
A^{(1)}&=C_F,\\
A^{(2)}&=\frac{C_F}{2}\left[C_A\left(\frac{67}{18}-\frac{\pi^2}{6}\right)-\frac{10}{9}T_Rn_f\right],\\
B^{(1)}&=-\frac{3}{2}C_F.
\end{align}

For the nonperturbative form factor $S_{\rm{NP}}$ in the $pp$ Drell-Yan process, a general parameterization associated with the TMD PDF has been proposed in Ref.~\cite{Sun:2014dqm} and it has the following form:
\begin{align}
S_{\mathrm{NP}}=g_1 b^2+g_2 \ln \frac{b}{b_*} \ln \frac{Q}{Q_0}+g_3 b^2\left(\left(x_0 / x_1\right)^\lambda+\left(x_0 / x_2\right)^\lambda\right).
\end{align}
In Ref.\cite{Sun:2014dqm}, the parameters $g_1$, $g_2$, and $g_3$ are fitted from the nucleon-nucleon Drell-Yan process data at the initial scale $Q_{0}^{2}=2.4 ~\mathrm{GeV}^{2}$ with $b_{\mathrm{max}}=1.5~\rm{GeV}^{-1}$, $x_0=0.01$, and $\lambda = 0.2$. With the extracted parameters, the fitted results are $g_1=0.212$, $g_2=0.84$, and $g_3=0$. 

Since the nonperturbative form factor $S_{\rm{NP}}$ for quarks and antiquarks satisfies the following relation~\cite{Prokudin:2015ysa}
\begin{align}
S_{\mathrm{NP}}^q(Q, b)+S_{\mathrm{NP}}^{\bar{q}}(Q, b)=S_{\mathrm{NP}}(Q, b),
\end{align}
and the quark and antiquark contributions to $S_{\rm{pert}}$ are the same:
\begin{align}
S_{\mathrm{pert}}^q\left(Q, b_*\right)=S_{\mathrm{pert}}^{\bar{q}}\left(Q, b_*\right)=S_{\mathrm{pert}}\left(Q, b_*\right) / 2,
\end{align}
the $S_{\rm{NP}}$ associated with the TMD distribution function of one of the initial protons can be expressed as
\begin{align}
S_{\mathrm{NP}}^{f_{q / p}}(Q, b)=\frac{g_1}{2} b^2+\frac{g_2}{2} \ln \frac{b}{b_*} \ln \frac{Q}{Q_0}.
\end{align}
With all the ingredients presented above, we can rewrite the $b$-space distribution function
\begin{align}
\tilde{f}_{ q / p}(x, b ; Q)=e^{-\frac{1}{2} S_{\mathrm{pert}}\left(Q, b_*\right)-S_{\mathrm{Np}}^{f_{q / p}}(Q, b)} \mathcal{F}\left(\alpha_s(Q)\right) \sum_i \int_{x}^{1}\frac{d \xi}{\xi} C_{q \leftarrow i}(\frac{x}{\xi},  \mu_{B}) f_{i/p}(\xi, \mu_{B}).
\label{eq.fip}
\end{align}

One should notice that the hard factors $\mathcal{F}\left(\alpha_s(Q)\right)$, which appear in Eq.~(\ref{eq.fip}) and $H_{UU}(Q;\mu)$ in Eq.~(\ref{eq.WUU}), are scheme dependent. In the two different schemes, the Collins-11 scheme~\cite{Collins:2011zzd} and the Ji-Ma-Yuan scheme~\cite{Ji:2004wu,Ji:2004xq}, the hard factors $\mathcal{F}\left(\alpha_s(Q)\right)$ and $H_{UU}(Q;\mu)$ have different forms:
\begin{align}
& \tilde{\mathcal{F}}^{\mathrm{JCC}}\left(\alpha_s(Q)\right)=1+\mathcal{O}\left(\alpha_s^2\right), \\
& \tilde{\mathcal{F}}^{\mathrm{JMY}}\left(\alpha_s(Q), \rho\right)=1+\frac{\alpha_s}{2 \pi} C_F\left[\ln \rho-\frac{1}{2} \ln ^2 \rho-\frac{\pi^2}{2}-2\right], \\
& H^{\mathrm{JCC}}(Q ; \mu)=1+\frac{\alpha_s(\mu)}{2 \pi} C_F\left(3 \ln \frac{Q^2}{\mu^2}-\ln ^2 \frac{Q^2}{\mu^2}+\pi^2-8\right), \\
& H^{\mathrm{JMY}}(Q ; \mu, \rho)=1+\frac{\alpha_s(\mu)}{2 \pi} C_F\left(\left(1+\ln \rho^2\right) \ln \frac{Q^2}{\mu^2}-\ln \rho^2+\ln ^2 \rho+2 \pi^2-4\right),
\end{align}
where $\rho$ is another parameter to regulate the light-cone singularity of TMD distributions.
However, if one absorbs both of the hard factors $H$ and $\mathcal{F}$ into the definition of the $C$-coefficients, the $C$-coefficients turn out to be $\rho$-independent (scheme-independent) and can be expressed as~\cite{Catani:2000vq}
\begin{align}
& C_{q \leftarrow q^{\prime}}^{\prime}\left(x, b ; \mu_b\right)=\delta_{q q^{\prime}}\left[\delta(1-x)+\frac{\alpha_s}{\pi}\left(\frac{C_F}{2}(1-x)+\frac{C_F}{4}\left(\pi^2-8\right) \delta(1-x)\right)\right], \label{eq.NEWC1}\\
& C_{q \leftarrow g}^{\prime}\left(x, b ; \mu_b\right)=\frac{\alpha_s}{\pi} T_R x(1-x) \label{eq.NEWC2}.
\end{align}
Substituting the above new $C$-coefficients from Eq.~(\ref{eq.NEWC1}) and Eq.~(\ref{eq.NEWC2}) into Eq.~(\ref{eq.collinear}), we can obtain 
\begin{align}
\tilde{f}_{q/p}(x, b ; \mu_{B})&= \sum_{i}C_{q \leftarrow i}^{\prime} \otimes f_{i/p}(x, \mu_{B})\nonumber\\
&=\left(1+\frac{\alpha_S}{\pi} \frac{C_F}{4}\left(\pi^2-8\right)\right) f_1^{q / p}\left(x, \mu_b\right)+\int_x^1 \frac{d \xi}{\xi} \frac{\alpha_S\left(\mu_b\right)}{\pi}\left(\frac{C_F}{2}\left(1-\frac{x}{\xi}\right)\right) f_1^{q / p}\left(\xi, \mu_b\right)\nonumber\\
&+\int_x^1 \frac{d \xi}{\xi} \frac{\alpha_S\left(\mu_b\right)}{\pi} T_R \frac{x}{\xi}\left(1-\frac{x}{\xi}\right) f_1^{g / p}\left(\xi, \mu_b\right). 
\end{align}

Finally, the structure function $\widetilde{W}_{U U}(Q ; b)$ in $b$-space can be written as 
\begin{align}
\widetilde{W}_{U U}(Q ; b) & =e^{-S_{\mathrm{per}}\left(Q^2, b\right)-S_{\mathrm{NP}}^{f_{1}^{q / p_{1}}}\left(Q^2, b\right)-S_{\mathrm{NP}}^{f_1^{q / p_{2}}}\left(Q^2, b\right)} \nonumber\\
& \times \sum_{q} e_q^2 C_{q \leftarrow i}^{\prime} \otimes f_{i / p_1}\left(x_1, \mu_b\right) C_{\bar{q} \leftarrow j}^{\prime} \otimes f_{j / p_2}\left(x_2, \mu_b\right) .
\label{eq.WUUresult}
\end{align}
Thus, substituting Eq.~(\ref{eq.WUUresult}) back into Eq.~(\ref{eq.CrossSection_TMD}), one can obtain the final expression for the differential cross section as
\begin{align}
\frac{d^4 \sigma}{d Q^2 d y d^2 \boldsymbol{q_{T}}}&=\sigma_0 \int \frac{d^{2} \boldsymbol{b}}{(2 \pi)^{2}} e^{i \boldsymbol{q_{T}} \cdot \boldsymbol{b}} \widetilde{W}_{U U}(Q ; b)\nonumber\\
&=\sigma_0 \int_0^{\infty} \frac{b d b}{2 \pi} J_0\left(q_{\perp} b\right) \times e^{-S_{\mathrm{per}}\left(Q^2, b\right)-S_{\mathrm{NP}}^{f_{1}^{q / p_{1}}}\left(Q^2, b\right)-S_{\mathrm{NP}}^{f_1^{q / p_{2}}}\left(Q^2, b\right)} \nonumber\\
& \times \sum_{q} e_q^2 C_{q \leftarrow i}^{\prime} \otimes f_{i / p1}\left(x_1, \mu_b\right) C_{\bar{q} \leftarrow j}^{\prime} \otimes f_{j / p2}\left(x_2, \mu_b\right).
\end{align}

\subsection{The differential cross section for the Drell-Yan process using the Gaussian Ansatz}
From Eq.~(\ref{eq.CrossSection}), it can be inferred that the differential cross section for the Drell-Yan process has the following form
\begin{align}
\frac{d^4 \sigma}{d Q^2 d y d^2 \boldsymbol{q_{T}}}=\sigma_0 \sum_{q} e_q^2 \int d^2\boldsymbol{k_{T1}}d^2 \boldsymbol{k_{T2}}f_1(x_1,Q,\boldsymbol{k_{T1}})f_1(x_2,Q,\boldsymbol{k_{T2}})\delta^2(\boldsymbol{q_{T}}-\boldsymbol{k_{T1}}-\boldsymbol{k_{T2}}).
\label{eq.gausssection}
\end{align}
The transverse momentum dependence of the unpolarized distribution functions  $f_1(x_1,Q,\boldsymbol{k_{T1}})$ and $f_1(x_2,Q,\boldsymbol{k_{T2}})$ can be assumed to follow a Gaussian-like distribution as
\begin{align}
f_1(x_1,Q,\boldsymbol{k_{T1}})=f_1(x_1,Q) \frac{1}{\pi \left\langle k_{T1}^2\right\rangle} \exp \left(-\frac{k_{T1}^2}{\left\langle k_{T1}^2\right\rangle}\right),\label{eq.gauss1}\\
f_1(x_2,Q,\boldsymbol{k_{T2}})=f_1(x_2,Q) \frac{1}{\pi \left\langle k_{T2}^2\right\rangle} \exp\left(-\frac{k_{T2}^2}{\left\langle k_{T2}^2\right\rangle}\right),\label{eq.gauss2}
\end{align}
where $\left\langle k_{T1}^2\right\rangle=\left\langle k_{T2}^2\right\rangle=\left\langle k_{T}^2\right\rangle=0.25~\rm{GeV}^2$ represents the mean value of the squared transverse momentum of the partons in the initial-state protons. Substituting Eq.~(\ref{eq.gauss1}) and Eq.~(\ref{eq.gauss2}) into Eq.~(\ref{eq.gausssection}), we can obtain
\begin{align}
\frac{d^4 \sigma}{d Q^2 d y d^2 \boldsymbol{q_{T}}}=&\sigma_0 \sum_{q} e_q^2 f_1(x_1,Q) \frac{1}{\pi \left\langle k_{T}^2\right\rangle}f_1(x_2,Q) \frac{1}{\pi \left\langle k_{T}^2\right\rangle}\int d^2\boldsymbol{k_{T1}}d^2 \boldsymbol{k_{T2}} \exp \left(-\frac{k_{T1}^2}{\left\langle k_{T}^2\right\rangle}\right) \exp \left(-\frac{k_{T2}^2}{\left\langle k_{T}^2\right\rangle}\right)\nonumber\\ 
&\delta^2(\boldsymbol{q_{T}}-\boldsymbol{k_{T1}}-\boldsymbol{k_{T2}}).
\end{align}
After further calculations, the following expression of the differential cross section is obtained
\begin{align}
\frac{d^4 \sigma}{d Q^2 d y d^2 \boldsymbol{q_{T}}}
=&\sigma_0 \sum_{q} e_q^2 f_1(x_1,Q) \frac{1}{\pi \left\langle k_{T}^2\right\rangle}f_1(x_2,Q) \frac{1}{\pi \left\langle k_{T}^2\right\rangle} \exp\left(\frac{-q_{T}^2}{2\left\langle k_{T}^2\right\rangle}\right) \frac{1}{2} \left\langle k_{T}^2\right\rangle \pi \nonumber\\
=&\frac{1}{2\pi \left\langle k_{T}^2\right\rangle}\exp\left(\frac{-q_{T}^2}{2\left\langle k_{T}^2\right\rangle}\right)\sigma_0 \sum_{q} e_q^2 f_1(x_1,Q) f_1(x_2,Q).
\end{align}

\section{NUMERICAL ESTIMATE}
\label{Sec.there}
In this section, we will perform the numerical calculation utilizing both methods and compare the results with the experimental data from the CMS Collaboration.

Regarding the CSS method, we adopt the NLO set of the CT10~\cite{Lai:2010vv} parametrization (central PDF sets) for the collinear unpolarized distribution function $f_{1}(x)$ of the proton as input. The values of the strong coupling $\alpha_{s}$ are consistently obtained in 2-loop order as
\begin{align}
\alpha_{s}\left(Q^{2}\right)=\frac{12 \pi}{\left(33-2 n_{f}\right) \ln \left(Q^{2} / \Lambda_{\rm{QCD}}^{2}\right)}\left\{1-\frac{6\left(153-19 n_{f}\right)}{\left(33-2 n_{f}\right)^{2}} \frac{\ln \ln \left(Q^{2} / \Lambda_{\rm{QCD}}^{2}\right)}{\ln \left(Q^{2} / \Lambda_{\rm{QCD}}^{2}\right)}\right\},
\label{eq.alpha}
\end{align}
with fixed $n_{f}=5$ and $\Lambda_{\rm{QCD}}=0.225 \rm{GeV}$. The running coupling in Eq.~(\ref{eq.alpha}) satisfies $\alpha_{s}(M_{Z}^{2}) = 0.118$.

\begin{figure}
\centering
\includegraphics[width=0.31\linewidth]{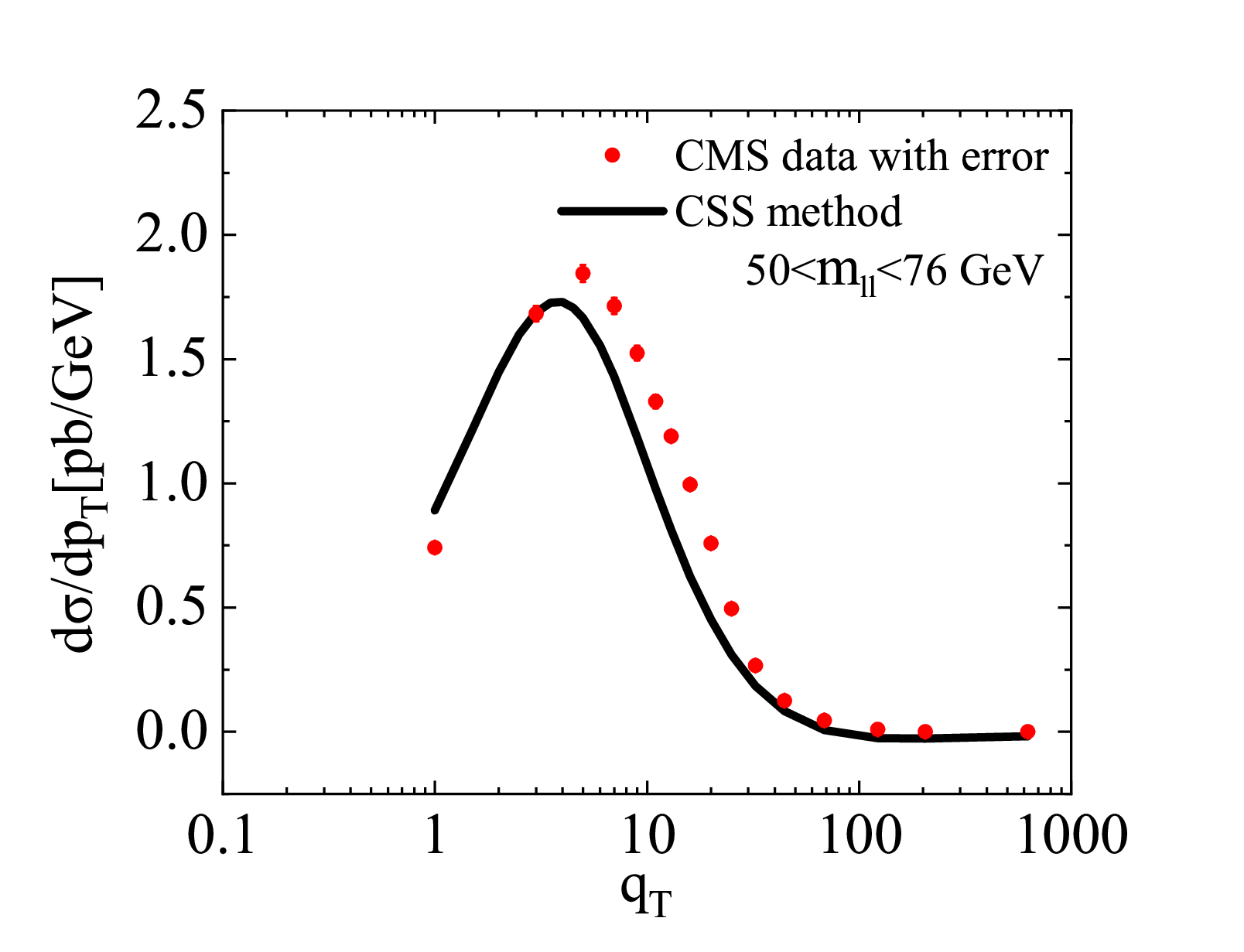}
\includegraphics[width=0.31\linewidth]{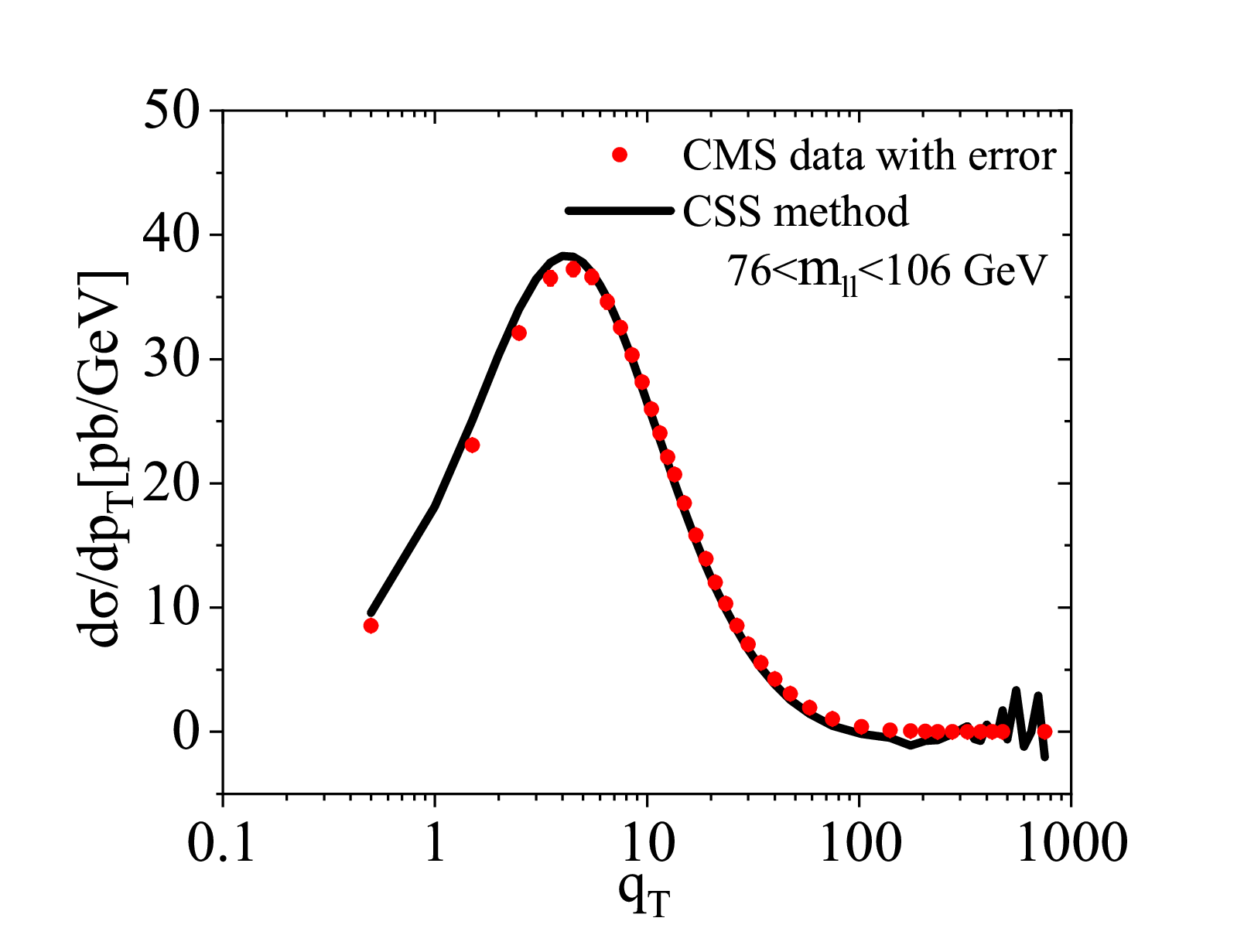}
\includegraphics[width=0.31\linewidth]{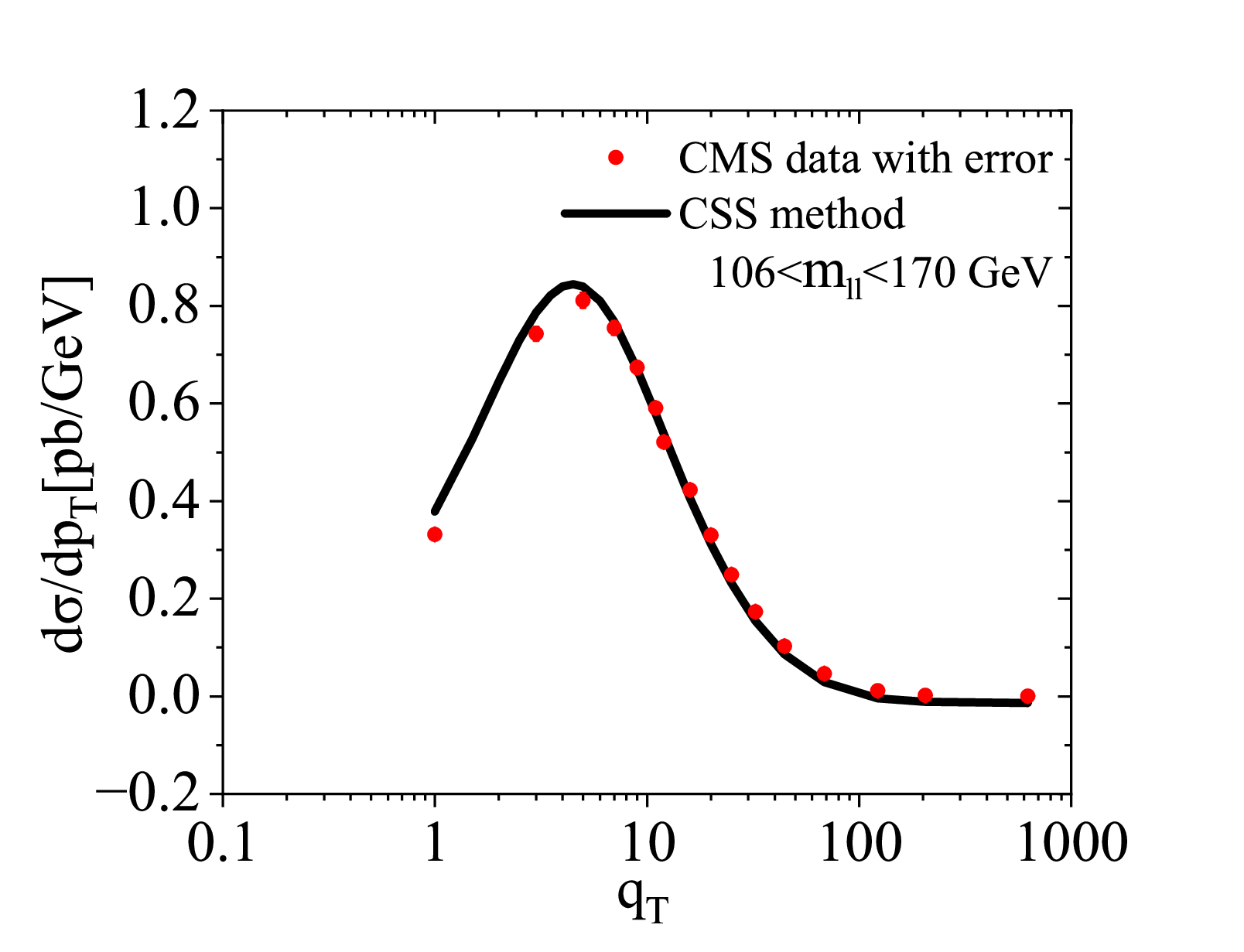}
\includegraphics[width=0.31\linewidth]{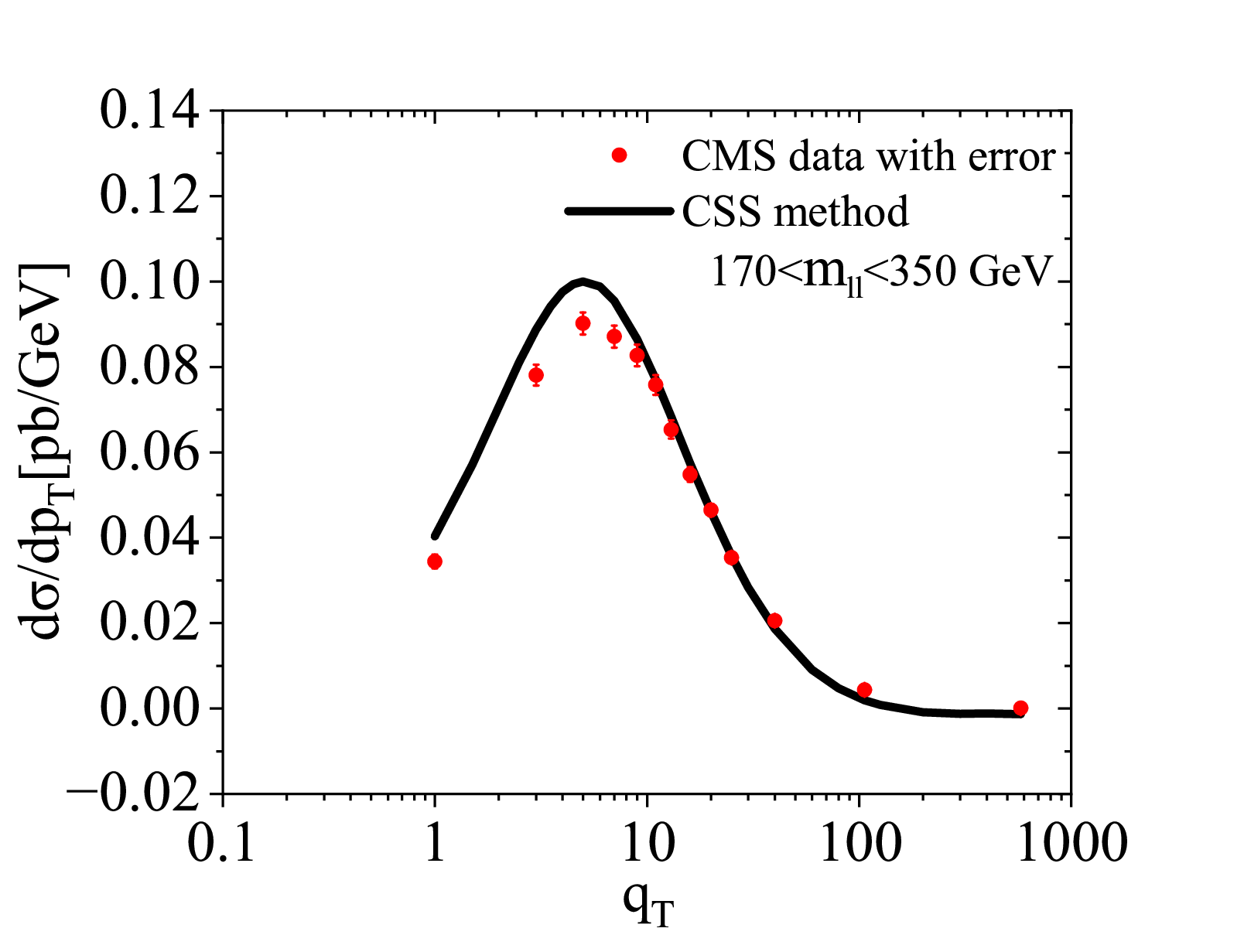}
\includegraphics[width=0.31\linewidth]{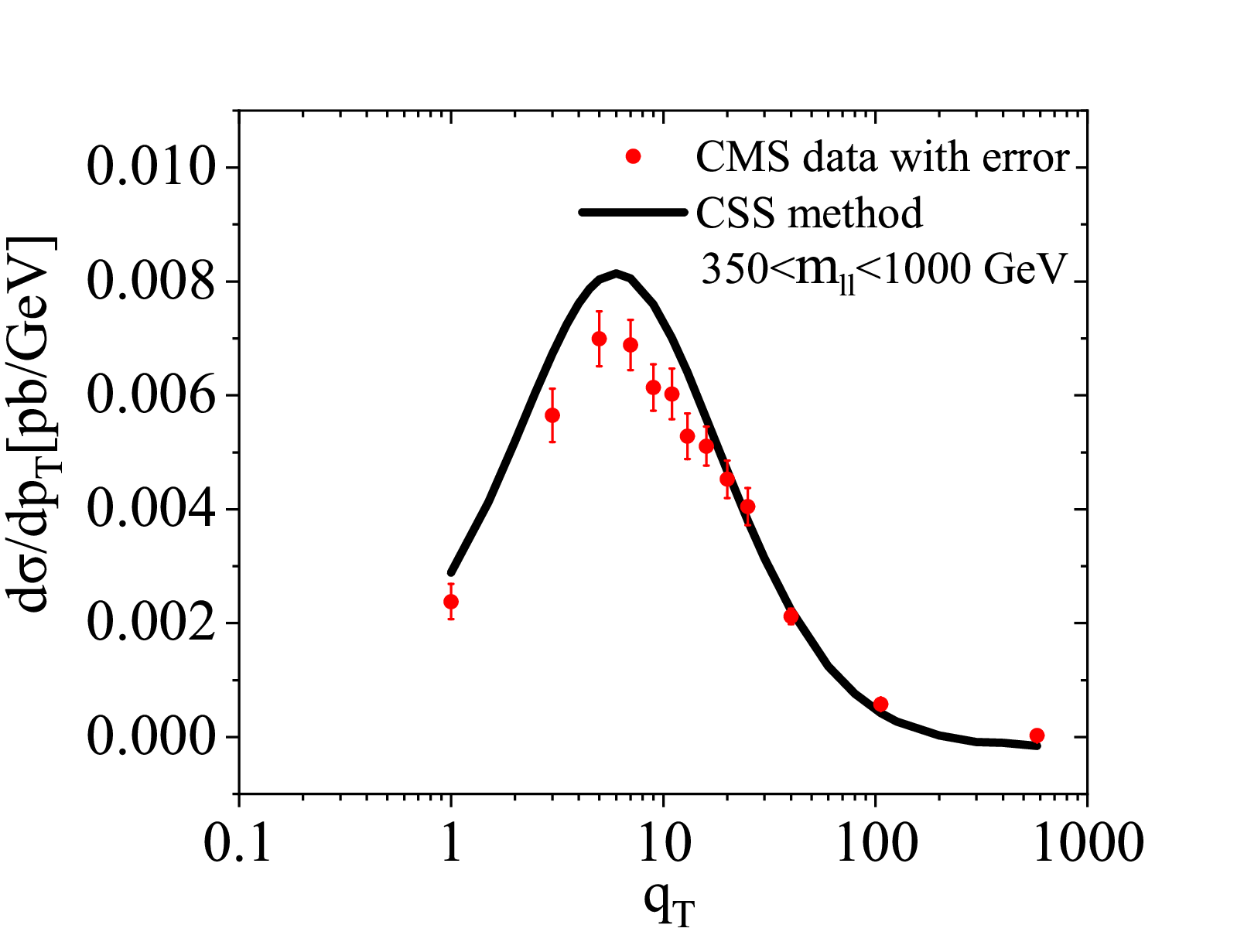}
\caption{The distribution of the differential cross section using the CSS method as a function of the lepton pair transverse momentum $q_{T}$ is presented for the lepton pair invariant mass ranges of $50~\rm{GeV}-76~\rm{GeV}$, $76~\rm{GeV}-106~\rm{GeV}$, $106~\rm{GeV}-170~\rm{GeV}$, $170~\rm{GeV}-350~\rm{GeV}$ and $350~\rm{GeV}-1000~\rm{GeV}$.}
\label{fig.Collins-Soper-Sterman}
\end{figure}

The CMS Collaboration has measured double-differential cross sections of Drell–Yan dilepton production as a function of the invariant mass and transverse momentum at $\sqrt{s}=13~\mathrm{TeV}$, using data collected in 2016, corresponding to an integrated luminosity of $36.3~\mathrm{fb}^{-1}$~\cite{CMS:2022ubq}.
It provides us with an ideal opportunity to compare the validity of two different methods for describing the TMD dependence of the PDFs.

The differential cross sections obtained from the two different calculation methods are both four-dimensional differential cross sections $\frac{d^4 \sigma}{d Q^2 d y d^2 \boldsymbol{q_{T}}}$ with respect to $Q^{2}$, $y$, and $\boldsymbol{q}_{T}^{2}$. However, the CMS Collaboration measured the cross section as a function of $q_{T}$ for different invariant mass bins: $\frac{d \sigma}{d  q_{T}}$. In order to be consistent with the experimental data, we integrate over the remaining two variables within the kinematical range of CMS and the corresponding invariant mass bins. The integration variables and their ranges corresponding to the kinematic coverage of the CMS data are as follows
\begin{align}
\frac{\mathrm{d} \sigma}{\mathrm{d} q_{T}}&=\sigma_0 q_{T} \int d Q^2 \int_{-2.4}^{+2.4} d y \int_0^{\infty} d b b J_0\left(q_{\perp} b\right) \times \widetilde{W}_{U U}(Q, b),\label{eq.css}\\
\frac{\mathrm{d} \sigma}{\mathrm{d} q_{T}}&=\sigma_0 q_{T} \int d Q^2 \int_{-2.4}^{2.4}d y  \frac{1}{ \left\langle k_{T}^2\right\rangle}exp\left(\frac{-q_{T}^2}{2\left\langle k_{T}^2\right\rangle}\right) \sum_{q} e_q^2 f_1(x_1,Q) f_1(x_2,Q).\label{eq.gauss}
\end{align}
As a comparison with experimental data, the following ranges of $Q$ have been adopted for the integration over $Q^{2}$: $50~\rm{GeV}-76~\rm{GeV}$, $76~\rm{GeV}-106~\rm{GeV}$, $106~\rm{GeV}-170~\rm{GeV}$, $170~\rm{GeV}-350~\rm{GeV}$, and $350~\rm{GeV}-1000~\rm{GeV}$. The integration range for $y$ is from -2.4 to 2.4. 

For dilepton invariant masses above the $Z$-boson mass, the contribution from $Z$-boson production becomes significant, and the above expressions must be modified accordingly.
Therefore, except for the first interval $Q^{2}=50~\rm{GeV}-76~\rm{GeV}$, we replace the quark electric charge $e_q$ with the electroweak charge $\hat{e}_q$ for the other intervals, as follows:
\begin{align}
\hat{e}_q^2&=e_q^2-2 e_q \Xi_1\left(Q^2\right) V_e V_q+\Xi_2\left(Q^2\right)\left(1+V_e^2\right)\left(1+V_q^2\right), \label{eq.Zeq}\\
\Xi_1(s)&=\frac{1}{16 \sin^2 \Theta_W \cos^2 \Theta_W} \frac{s\left(s-M_Z^2\right)}{\left(s-M_Z^2\right)^2+\Gamma_Z^2 M_Z^2}, \\
\Xi_2(s)&=\frac{1}{256 \sin^4 \Theta_W \cos^4 \Theta_W} \frac{s^2}{\left(s-M_Z^2\right)^2+\Gamma_Z^2 M_Z^2}.
\end{align}
The electroweak couplings are given in terms of the Weinberg angle $\Theta_W$ by
\begin{align}
V_e & = -1 + 4 \sin^2 \Theta_W, \\
V_u & = +1 - \frac{8}{3} \sin^2 \Theta_W, \\
V_d & = -1 + \frac{4}{3} \sin^2 \Theta_W,
\end{align}
with $\sin^2 \Theta_W (\rm{MS}) = 0.23122$, $M_Z = 91.1876~\rm{GeV}$, and $\Gamma_Z = 2.4952~\rm{GeV}$ from the PDG~\cite{ParticleDataGroup:2020ssz}. The couplings $V_u$ and $V_d$ are for up and down type quarks, respectively.

The numerical results are shown in Fig.~\ref{fig.Collins-Soper-Sterman}, where the red data points represent the experimental measurements with errors, and the solid black line corresponds to the theoretical calculations obtained using the CSS method. 
From Fig.~\ref{fig.Collins-Soper-Sterman}, we can see that the theoretical predictions based on the CSS resummation describe the general shape of the data well over wide ranges of $q_T$ and $Q$. The comparison demonstrates that the CSS resummation framework captures the essential features of the $q_T$ spectrum across a broad range of invariant masses.

\begin{figure}
\centering
\includegraphics[width=0.31\linewidth]{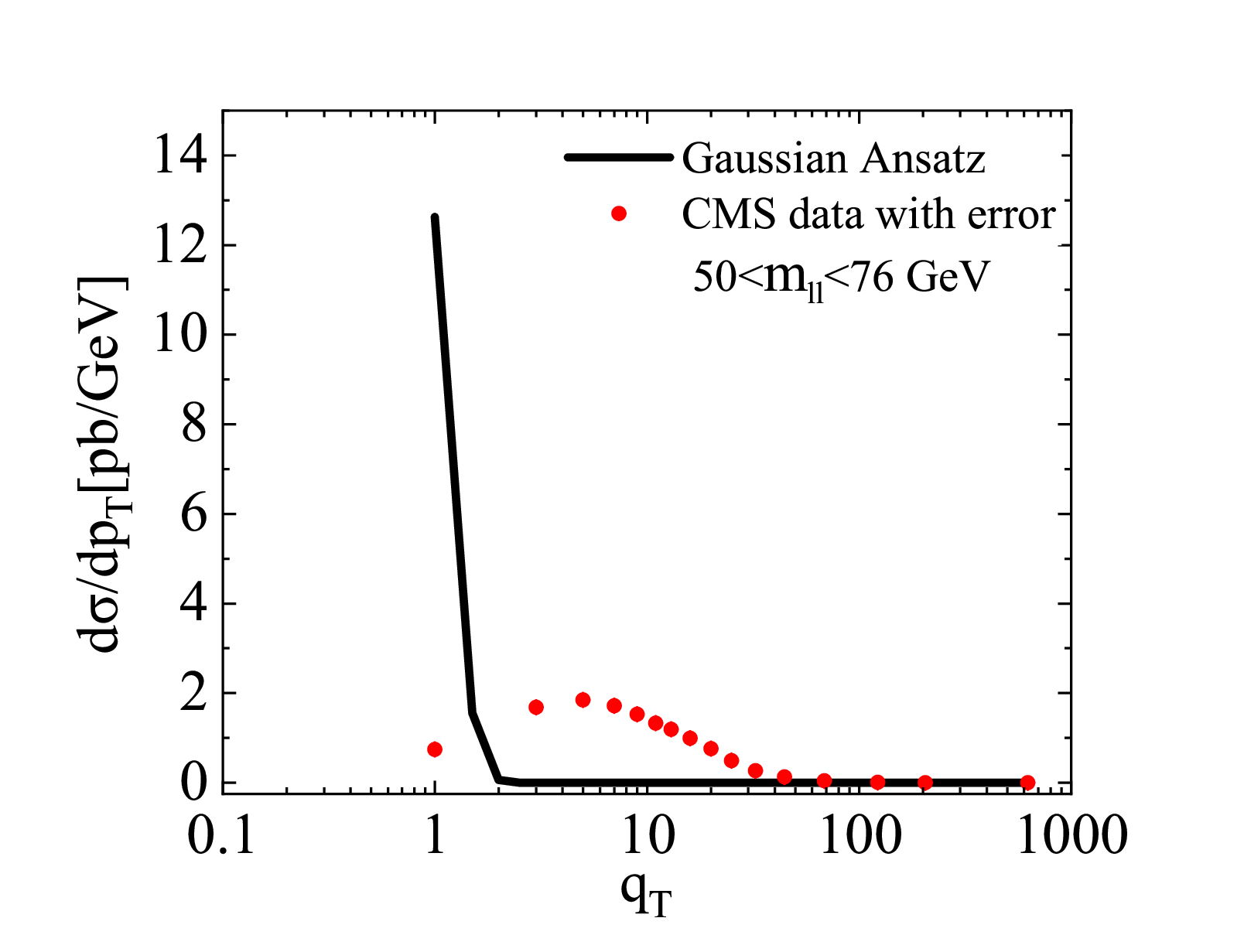}
\includegraphics[width=0.31\linewidth]{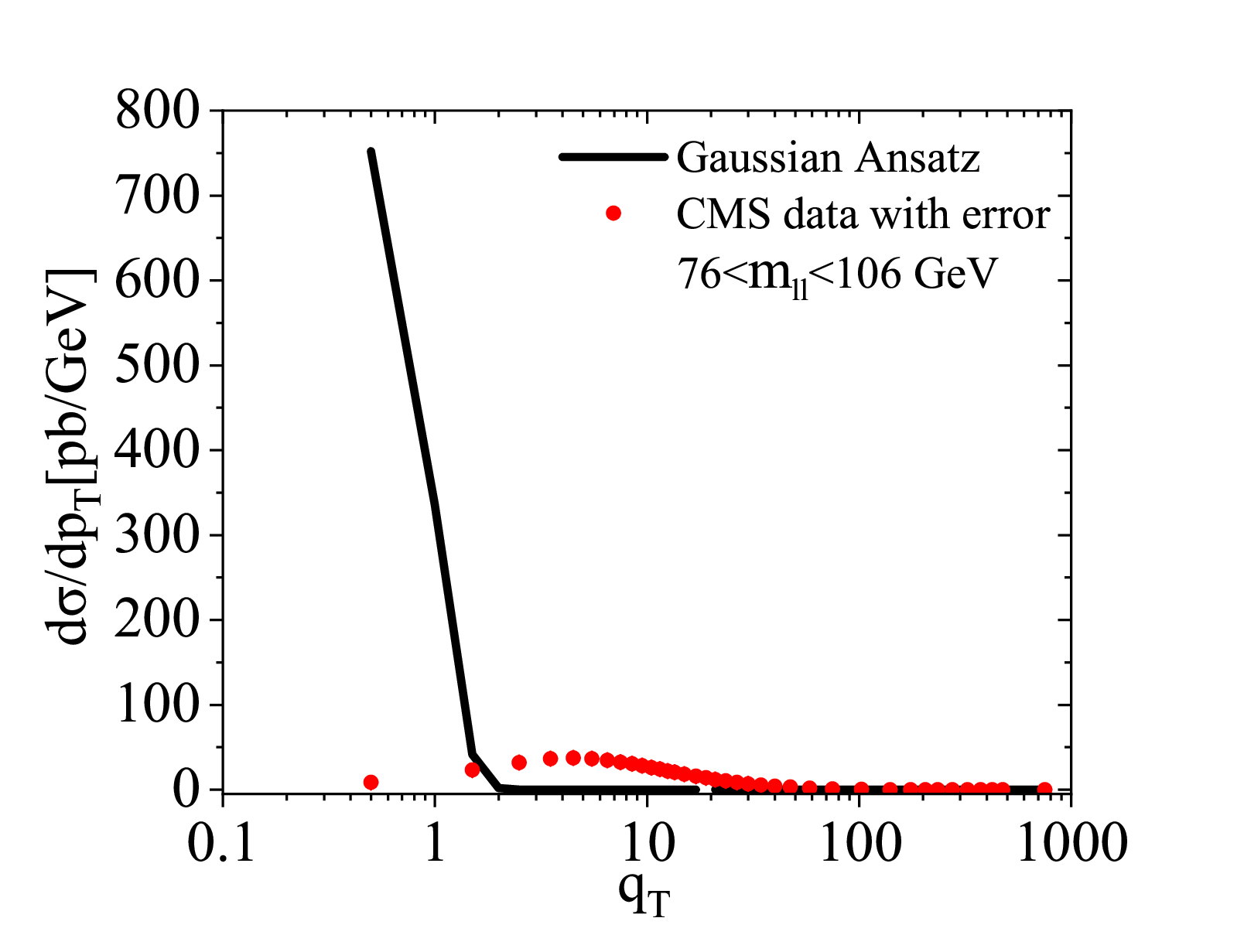}
\includegraphics[width=0.31\linewidth]{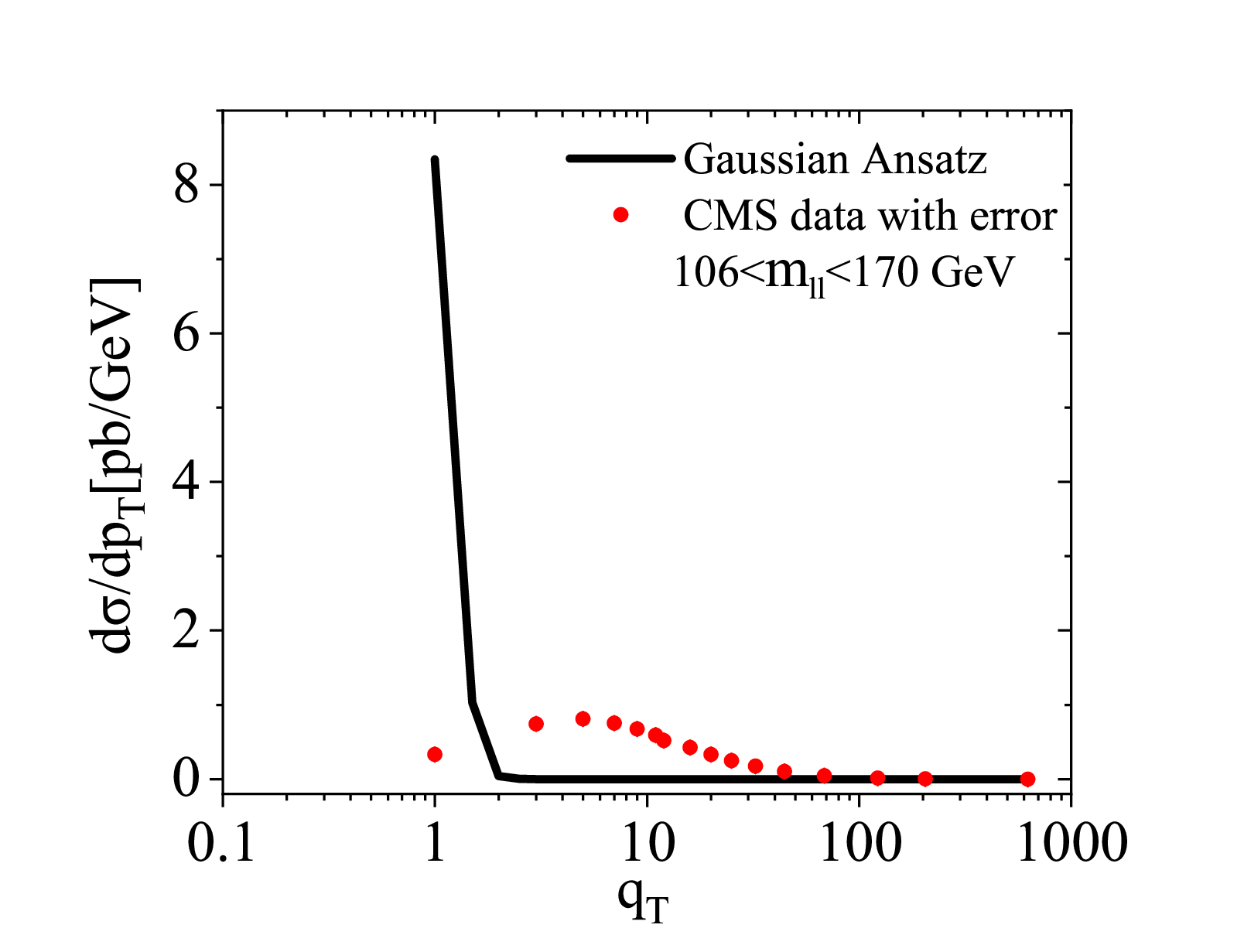}
\includegraphics[width=0.31\linewidth]{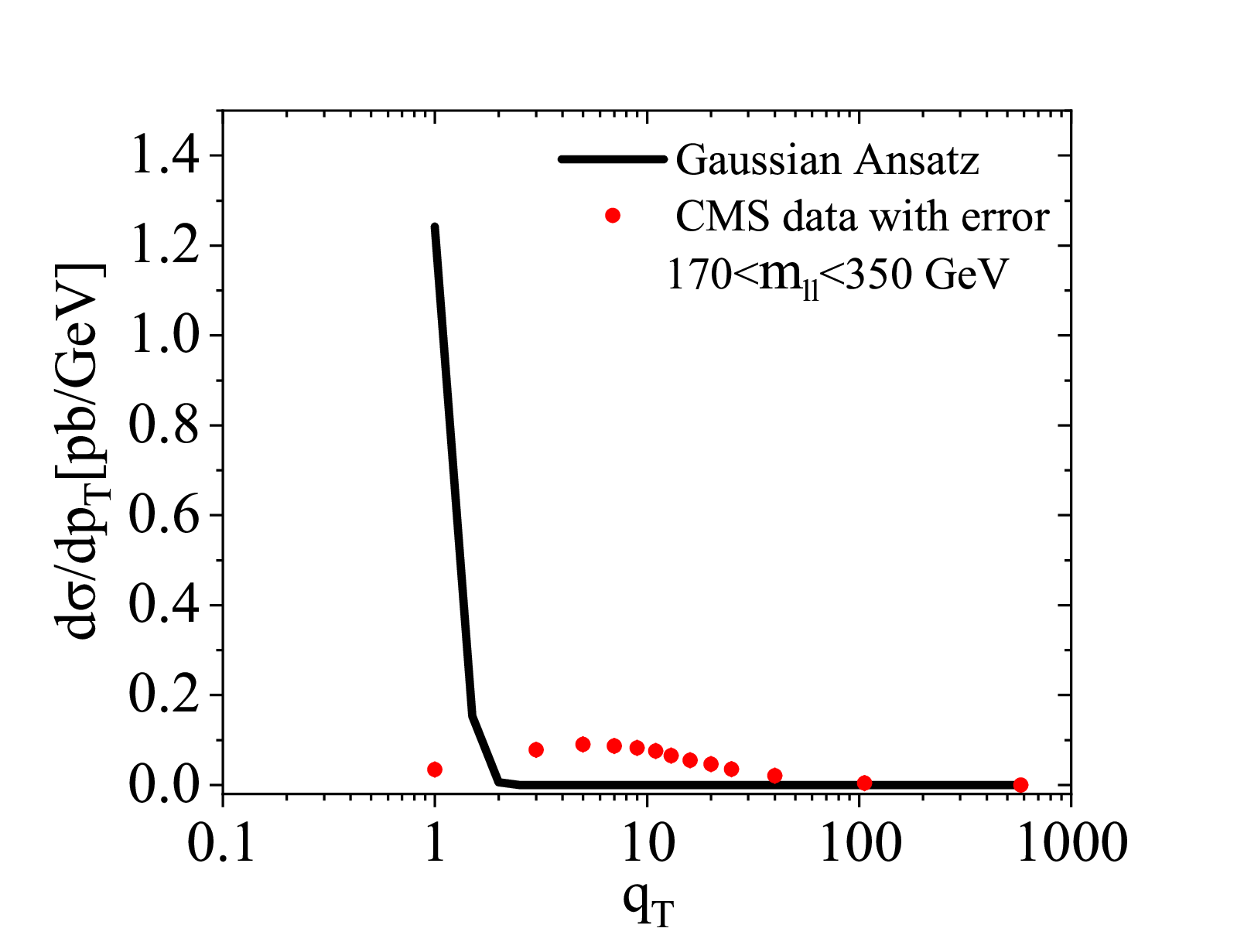}
\includegraphics[width=0.31\linewidth]{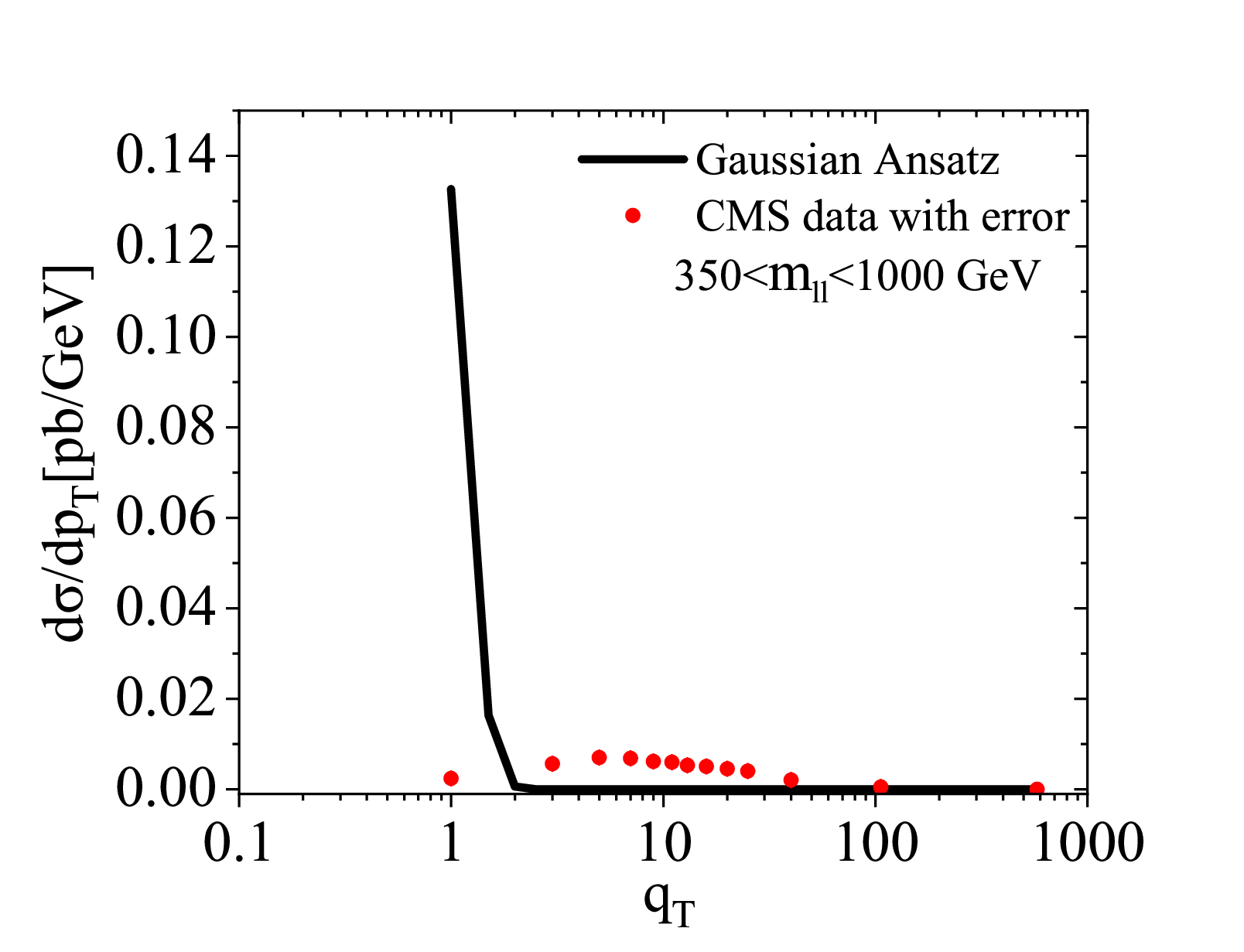}
\caption{The distribution of the differential cross section using the Gaussian Ansatz as a function of the lepton pair transverse momentum $q_{T}$ is presented for the lepton pair invariant mass ranges of $50~\rm{GeV}-76~\rm{GeV}$, $76~\rm{GeV}-106~\rm{GeV}$, $106~\rm{GeV}-170~\rm{GeV}$, $170~\rm{GeV}-350~\rm{GeV}$ and $350~\rm{GeV}-1000~\rm{GeV}$.}
\label{fig.Gaussian}
\end{figure}

\begin{figure}
\centering
\includegraphics[width=0.45\linewidth]{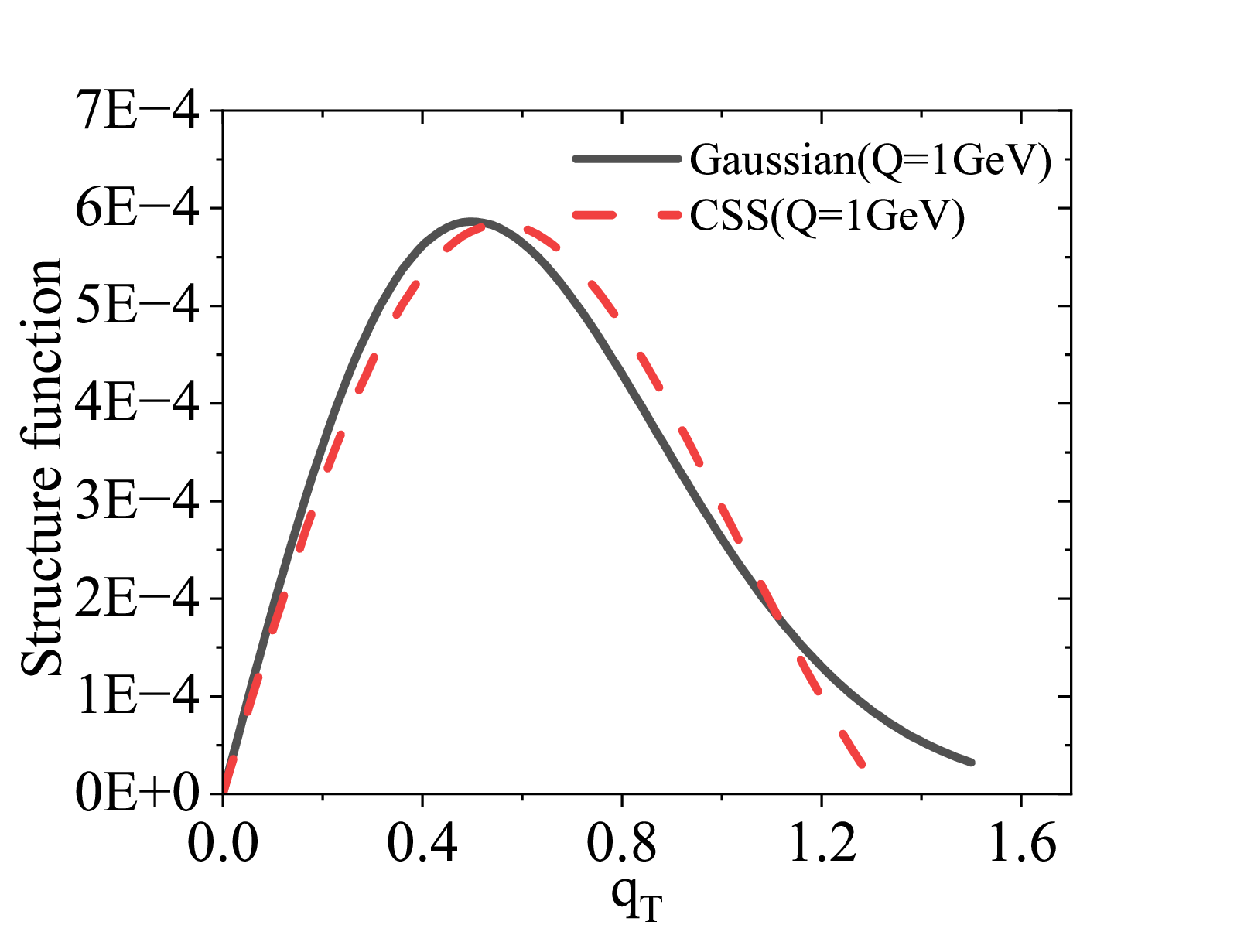}
\includegraphics[width=0.45\linewidth]{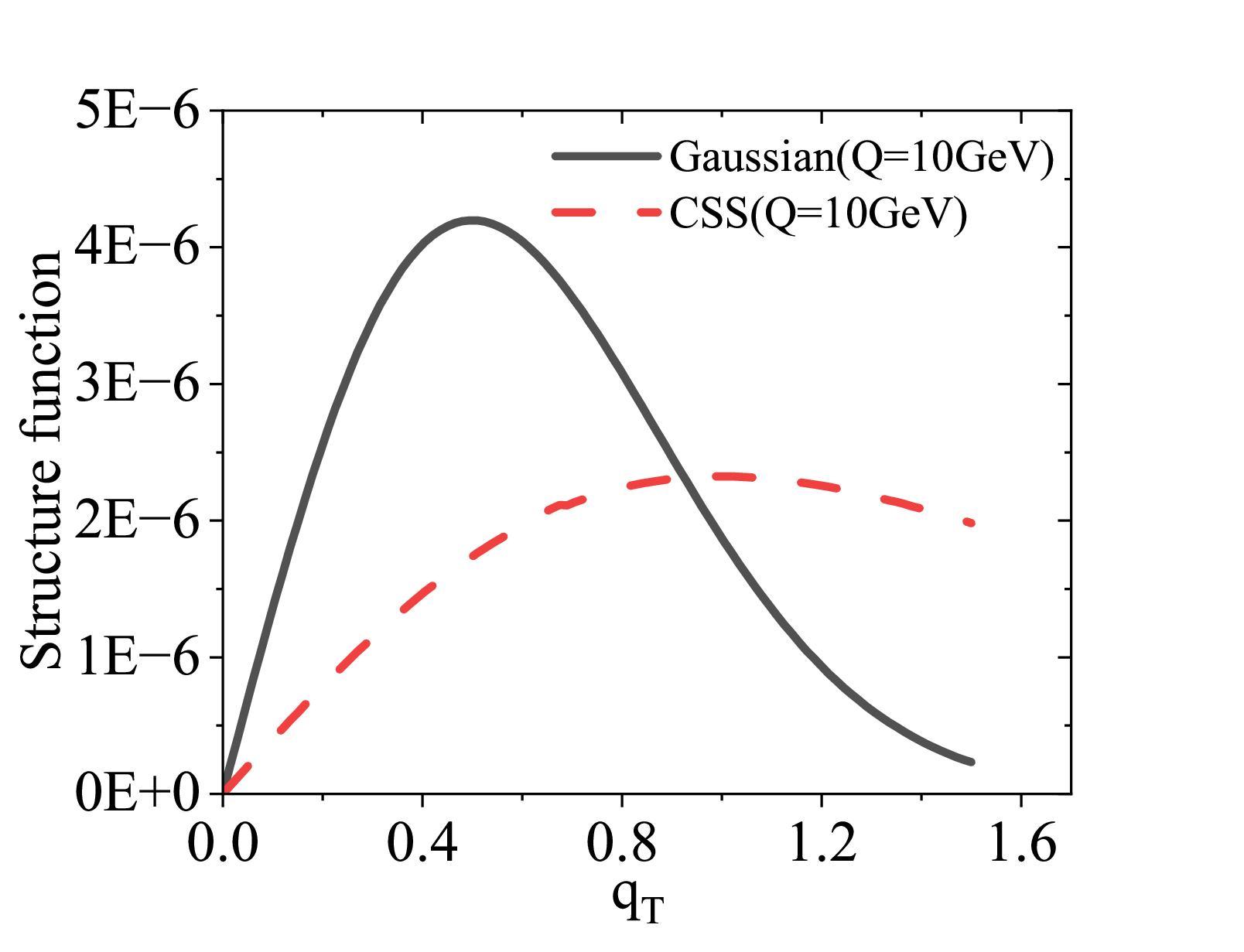}
\includegraphics[width=0.45\linewidth]{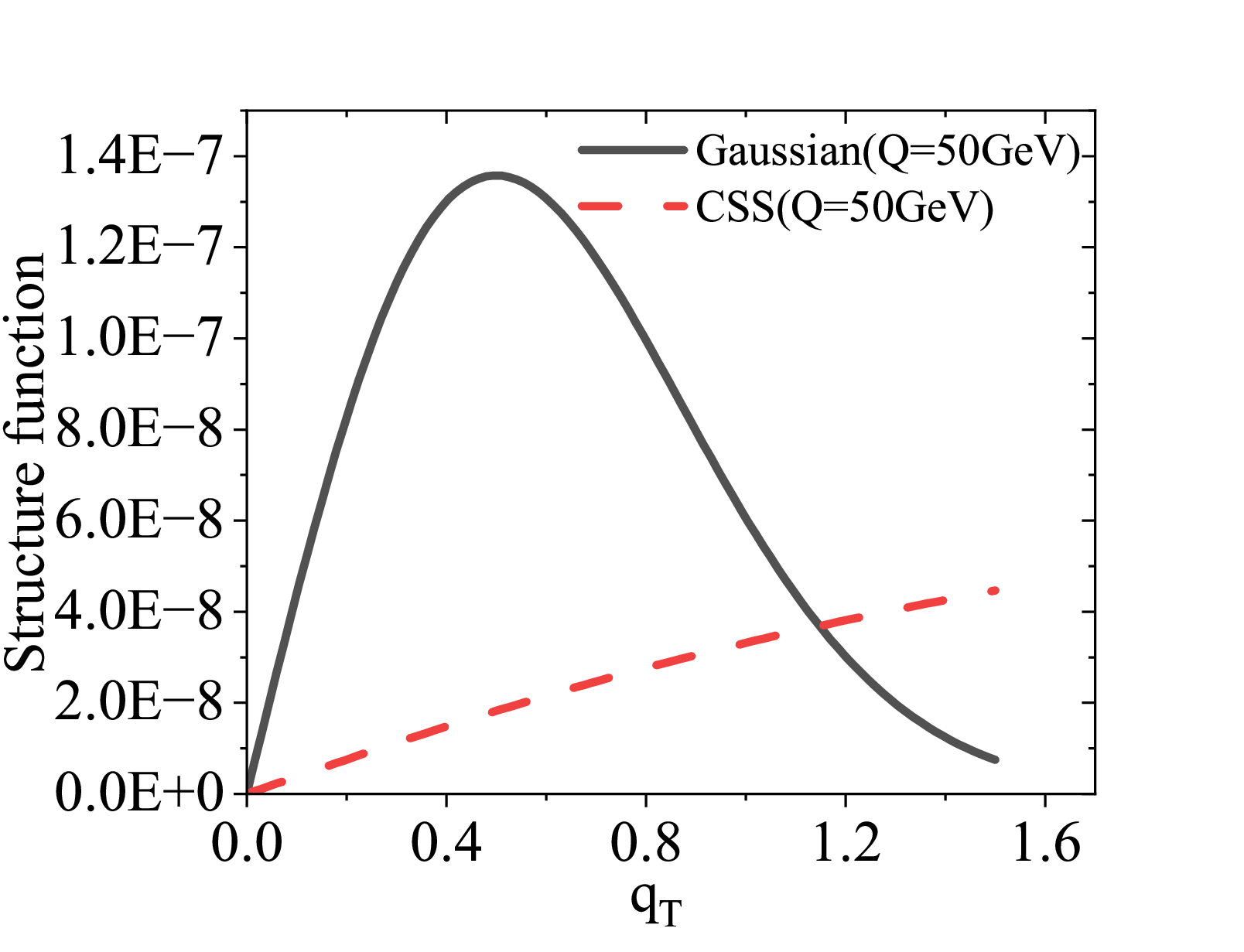}
\includegraphics[width=0.45\linewidth]{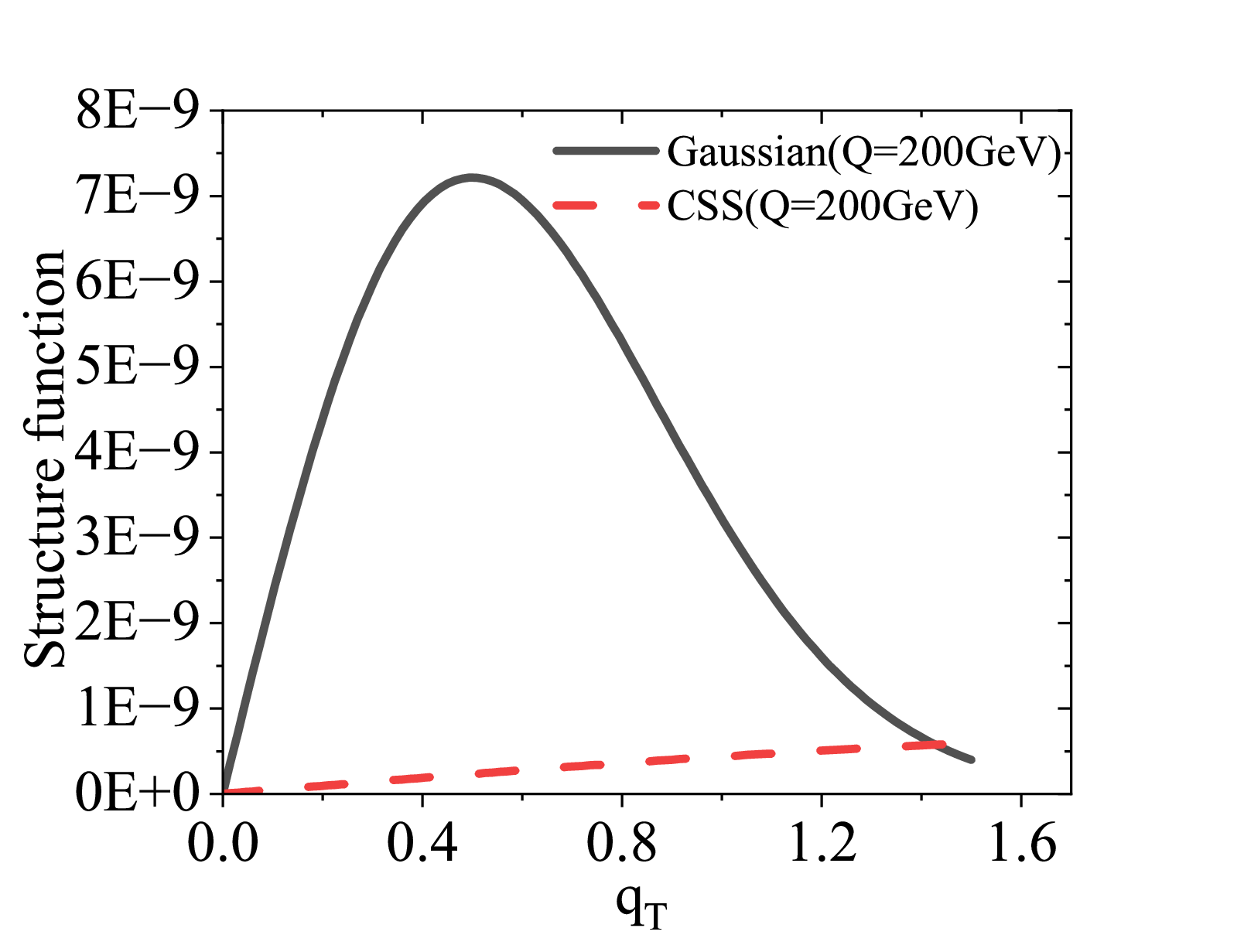}
\caption{Setting $x_{1}=0.1$ and $x_{2}=0.2$, the distribution of the structure function as a function of $q_{T}$ was calculated for $Q=1~\rm {GeV}$, $Q=10~\rm {GeV}$, $Q=50~\rm {GeV}$ and $Q=200~\rm {GeV}$. The red solid line represents the distribution under the CSS approach, while the black solid line corresponds to the result from the Gaussian method.}
\label{fig.daussia}
\end{figure}

The numerical estimate of the differential cross section from the Gaussian Ansatz with the experimental measurement is shown in Fig.~\ref{fig.Gaussian}. From the figure, one can conclude that the Gaussian Ansatz of the TMD PDFs cannot describe the experimental data at high energy scales. 
The results of the structure functions utilizing the CSS method and the Gaussian Ansatz are shown in Fig.~\ref{fig.daussia}.
The figure shows the structure function of the CSS and the structure function of the Gaussian Ansatz as a function of the lepton pair transverse momentum $q_T$ at $x_1 = 0.1$ and $x_2 = 0.2$, for energies $Q = 1\text{ GeV}$, $Q = 10\text{ GeV}$, $Q = 50\text{ GeV}$, and $Q = 200\text{ GeV}$. The structure functions are as follows:
\begin{align}
F_{\rm{CSS}} &= \sigma_0 q_{T} \int_0^{\infty} b db J_0\left(q_{\perp} b\right) \times \widetilde{W}_{U U}(Q, b), \\
F_{\rm{Gaussian}} &= \sigma_0 q_{T} \frac{1}{ \left\langle k_{T}^2\right\rangle}exp\left(\frac{-q_{T}^2}{2\left\langle k_{T}^2\right\rangle}\right) \sum_{q} e_q^2 f_1(x_1,Q) f_1(x_2,Q).
\end{align}

In the figure, the dashed line represents the structure function calculated using the CSS method, while the solid line corresponds to the results from the Gaussian method. It can be observed that at low energies, the results from the Gaussian method agree well with those from the CSS method. However, as the energy increases, the Gaussian method begins to deviate from the CSS results and decreases rapidly. Therefore, it can be shown that the Gaussian method is only compatible with the CSS results in low-energy scenarios, which can describe the experimental data well over a wide energy range.
Therefore, we conclude that the CSS formalism provides a reliable description of the TMD dependence of TMD functions across a broad range of energy scales, while the Gaussian Ansatz breaks down at high energy scales.

\section{conclusion}
\label{Sec.four}
In this work, we have studied the transverse momentum dependent differential cross section in the $pp$ Drell-Yan process and compared our predictions with the CMS data collected in 2016, with an integrated luminosity of 36.3 fb$^{-1}$. Within the TMD factorization framework, the calculation utilized two distinct ways to deal with the transverse momentum dependence of TMD PDFs.
The first strategy is the Collins-Soper-Sterman (CSS) resummation formalism, where the evolution from an initial scale $\mu_B$ to a hard scale $Q$ is encoded in a Sudakov form factor, $\exp[-S(Q,b)]$. The second strategy is a Gaussian-type ansatz for the transverse-momentum dependence. 

The numerical comparisons show that the CSS resummation formalism provides a consistent description of the measured $q_T$ spectra over a wide range of invariant masses, while the Gaussian Ansatz is only compatible with CSS expectations at relatively low $Q$ and breaks down at higher scales.
These results support the applicability and predictive power of TMD factorization with CSS evolution for collider Drell-Yan observables, which help us better understand the scope of applicability and predictability of the TMD factorization theorem. In the future, theoretical calculations based on TMD factorization with CSS evolution can be employed to predict or validate the measurements from collider experiments to further constrain the three-dimensional partonic structure of the nucleon.

\section*{Acknowledgments}
This work is supported by the Natural Science Foundation of Henan Province under Grant Nos. 242300421377 and 232300421140.

\end{document}